\documentstyle[12pt,epsf]{article}
\newcommand{\nc}{\newcommand}
\nc{\renc}{\renewcommand}
\renc{\baselinestretch}{1.1}
\nc{\com}[1]{\ \\ \ {\bf \# {#1}}\\ \ }

\nc{\bort}[1]{}
\newlength{\overeqskip}
\newlength{\undereqskip}
\nc{\be}[1]{\begin{equation} \mbox{$\label{#1}$}}
\nc{\bea}[1]{\begin{eqnarray} \mbox{$\label{#1}$}}
\nc{\Section}[2]{\section{\sc #2}\label{#1}\seqnoll}
\nc{\Subsection}[2]{\subsection{\sc #2}\label{#1}}
\nc{\Bibitem}[1]{\bibitem{#1}}
\nc{\Label}[1]{\label{#1}}

\nc{\eea}{\vspace{\undereqskip}\end{eqnarray}}
\nc{\ee}{\vspace{\undereqskip}\end{equation}}
\nc{\bdm}{\begin{displaymath}}
\nc{\edm}{\end{displaymath}}
\nc{\dpsty}{\displaystyle}
\nc{\bc}{\begin{center}}
\nc{\ec}{\end{center}}
\nc{\ba}{\begin{array}}
\nc{\ea}{\end{array}}
\nc{\bab}{\begin{abstract}}
\nc{\eab}{\end{abstract}}
\nc{\btab}{\begin{tabular}}
\nc{\etab}{\end{tabular}}
\nc{\bit}{\begin{itemize}}
\nc{\eit}{\end{itemize}}
\nc{\ben}{\begin{enumerate}}
\nc{\een}{\end{enumerate}}
\nc{\bfig}{\begin{figure}}
\nc{\efig}{\end{figure}}
\nc{\seqnoll}{\setcounter{equation}{0}}
\renc{\theequation}{\thesection.\arabic{equation}}

\nc{\refc}[1]{\mbox{Ref.~\cite{#1}}}
\nc{\refs}[1]{\mbox{Refs.~\cite{#1}}}

\nc{\eqs}[2]{\mbox{Eqs.~(\ref{#1}) and (\ref{#2})}}
\nc{\eq}[1]{\mbox{Eq.~(\ref{#1})}}

\nc{\figs}[2]{\mbox{Figs.~\ref{#1} and \ref{#2}}}
\nc{\fig}[1]{\mbox{Fig.~\ref{#1}}}

\nc{\figcap}[1]{\begin{quote}\refstepcounter{figure}
        {\bf Figure \thefigure}: {\small\sl #1}\end{quote}}
\nc{\tabcap}[1]{\refstepcounter{table}
        {\bf Table \thetable}: {\small\sl #1}}

\nc{\tag}[1]{\label{#1} \marginpar{{\footnotesize #1}}}
\nc{\mtag}[1]{\label{#1} \mbox{\marginpar{{\footnotesize #1}}}}

\nc{\arreq}{&\!\!\!=\!\!\!&}
\nc{\arrmi}{&\!\!\!!-\!\!\!&}
\nc{\arrpl}{&\!\!\!+\!\!\!&}
\nc{\arrap}{&\!\!\!\approx\!\!\!&}
\nc{\non}{\nonumber}
\nc{\nn}{\nonumber\\}
\nc{\align}{\!\!\!\!\!\!\!\!&&}

\nc{\mat}[4]{{\left(\ba{cc} #1 & #2 \\ #3 & #4 \ea\right)}}

\def\simleq{\; \raise0.3ex\hbox{$<$\kern-0.75em
      \raise-1.1ex\hbox{$\sim$}}\; }
\def\simgeq{\; \raise0.3ex\hbox{$>$\kern-0.75em
      \raise-1.1ex\hbox{$\sim$}}\; }

\def\lsim{\simleq}
\nc{\DOT}{\hspace{-0.08in}{\bf .}\hspace{0.1in}}
\nc{\Laada}{\hbox {$\sqcap$ \kern -1em $\sqcup$}}
\nc\loota{{\scriptstyle\sqcap\kern-0.55em\hbox{$\scriptstyle\sqcup$}}}
\nc\Loota{{\sqcap\kern-0.65em\hbox{$\sqcup$}}}
\nc\laada{\Loota}
\nc{\qed}{\hskip 3em \hbox{\BOX} \vskip 2ex}
\def\Re{{\rm Re}\hskip2pt}
\def\Im{{\rm Im}\hskip2pt}
\nc{\real}{{\rm I \! R}}
\nc{\Z}{{\sf Z \!\!\! Z}}
\nc{\complex}{{\rm C\!\!\! {\sf I}\,\,}}

\def\bigid{\leavevmode\hbox{\small1\kern-3.8pt\normalsize1}}
\def\id{\leavevmode\hbox{\small1\kern-3.3pt\normalsize1}}
\nc{\slask}{\hspace{0.em}\not\hspace{-0.25em}}
\nc{\bis}{{\prime\prime}}
\nc{\pa}{\partial}
\nc{\na}{\nabla}
\def\>{\rangle}
\def\<{\langle}
\nc{\goto}{\rightarrow}
\nc{\swap}{\leftrightarrow}

\nc{\EE}[1]{ \mbox{$\times 10^{#1}$} }
\nc{\abs}[1]{\left|#1\right|}
\nc{\at}[2]{\left.#1\right|_{#2}}
\nc{\norm}[1]{\|#1\|}
\nc{\abscut}[2]{\abs{#1}_{\scriptscriptstyle#2}}
\nc{\vek}[1]{\vec{#1}}
\nc{\integral}[2]{\int\limits_{#1}^{#2}}
\nc{\inv}[1]{\frac{1}{#1}}
\nc{\dd}[2]{{{\partial #1}\over{\partial #2}}}
\nc{\ddd}[2]{{{{\partial}^2 #1}\over{\partial {#2}^2}}}
\nc{\dddd}[3]{{{{\partial}^2 #1}\over{\partial #2 \partial #3}}}
\nc{\dder}[2]{{{d #1}\over{d #2}}}
\nc{\ddder}[2]{{{d^2 #1}\over{d {#2}^2}}}
\nc{\dddder}[3]{{d^2 #1}\over{d #2 d #3}}
\nc{\dx}[1]{d\,^{#1}x}
\nc{\dy}[1]{d\,^{#1}y}
\nc{\dz}[1]{d\,^{#1}z}
\nc{\dbar}[2]{\frac{d\,^{#1}#2}{(2\pi)^{#1}}}

\nc{\cc}{\mbox{$c.c.$ }}
\nc{\hc}{\mbox{$h.c.$ }}
\nc{\cf}{cf.\ }
\nc{\erfc}{{\rm erfc}}
\nc{\Tr}{{\rm Tr\,}}
\nc{\tr}{{\rm tr\,}}
\nc{\pol}{{\rm pol}}
\nc{\sign}{{\epsilon}}
\nc{\bfT}{{\bf T }}

\def\MeV{{\rm\ MeV}}

\nc{\cA}{{\cal A}}
\nc{\cB}{{\cal B}}
\nc{\cD}{{\cal D}}
\nc{\cE}{{\cal E}}
\nc{\cF}{{\cal F}}
\nc{\cG}{{\cal G}}
\nc{\cH}{{\cal H}}
\nc{\cL}{{\cal L}}
\nc{\cM}{{\cal M}}
\nc{\cO}{{\cal O}}
\nc{\cP}{{\cal P}}
\nc{\cQ}{{\cal Q}}
\nc{\cT}{{\cal T}}

\nc{\etal}{\mbox{\it et al. }}
\nc{\ie}{{\rm i.e. }}
\nc{\eg}{{\it e.g. }}
\nc{\1}{\aa}
\nc{\2}{\"{a}}
\nc{\3}{\"{o}}
\nc{\4}{\AA}
\nc{\5}{\"{A}}
\nc{\6}{\"{O}}
\nc{\al}{\alpha}
\nc{\Del}{\Delta}
\nc{\e}{\epsilon}
\nc{\eps}{\epsilon}
\nc{\g}{\gamma}
\nc{\lam}{\lambda}
\nc{\om}{\omega}
\nc{\Om}{\Omega}
\nc{\ve}{\varepsilon}
\nc{\mn}{{\mu\nu}}
\nc{\ka}{\kappa}

\nc{\vp}{\varphi}
\nc{\pub}[4]{\Bibitem{#1}#2, {\sl ``#3''}, #4.}
\nc{\advp}[3]{{\it  Adv.\ in\ Phys.\ }{{\bf #1} {(#2)} {#3}}}
\nc{\annp}[3]{{\it  Ann.\ Phys.\ (N.Y.)\ }{{\bf #1} {(#2)} {#3}}}
\nc{\apl}[3]{{\it  Appl. Phys. Lett. }{{\bf #1} {(#2)} {#3}}}
\nc{\apj}[3]{{\it  Astrophys.\ J.\ }{{\bf #1} {(#2)} {#3}}}
\nc{\apjl}[3]{{\it  Astrophys.\ J.\ Lett.\ }{{\bf #1} {(#2)} {#3}}}
\nc{\app}[3]{{\it Astropart.\ Phys.\ }{{\bf #1} {(#2)} {#3}}}

\nc{\cmp}[3]{{\it  Comm.\ Math.\ Phys.\ }{{ \bf #1} {(#2)} {#3}}}
\nc{\cqg}[3]{{\it  Class.\ Quant.\ Grav.\ }{{\bf #1} {(#2)} {#3}}}
\nc{\epl}[3]{{\it  Europhys.\ Lett.\ }{{\bf #1} {(#2)} {#3}}}
\nc{\ijmp}[3]{{\it Int.\ J.\ Mod.\ Phys.\ }{{\bf #1} {(#2)} {#3}}}
\nc{\ijtp}[3]{{\it Int.\ J.\ Theor.\ Phys.\ }{{\bf #1} {(#2)} {#3}}}
\nc{\jmp}[3]{{\it  J.\ Math.\ Phys.\ }{{ \bf #1} {(#2)} {#3}}}
\nc{\jpa}[3]{{\it  J.\ Phys.\ A\ }{{\bf #1} {(#2)} {#3}}}
\nc{\jpc}[3]{{\it  J.\ Phys.\ C\ }{{\bf #1} {(#2)} {#3}}}
\nc{\jpg}[3]{{\it J.~Phys.~G:~Nucl.~Part.~Phys.~}{{\bf #1} {(#2)} {#3}}}
\nc{\jap}[3]{{\it J.\ Appl.\ Phys.\ }{{\bf #1} {(#2)} {#3}}}
\nc{\jpsj}[3]{{\it J.\ Phys.\ Soc.\ Japan\ }{{\bf #1} {(#2)} {#3}}}
\nc{\lmp}[3]{{\it Lett.\ Math.\ Phys.\ }{{\bf #1} {(#2)} {#3}}}
\nc{\lncim}[3]{{\it  Lett.\ Nuov.\ Cim.\ }{{\bf #1} {(#2)} {#3}}}
\nc{\mpl}[3]{{\it  Mod.\ Phys.\ Lett.\ }{{\bf #1} {(#2)} {#3}}}
\nc{\ncim}[3]{{\it  Nuov.\ Cim.\ }{{\bf #1} {(#2)} {#3}}}
\nc{\np}[3]{{\it  Nucl.\ Phys.\ }{{\bf #1} {(#2)} {#3}}}
\nc{\pr}[3]{{\it Phys.\ Rev.\ }{{\bf #1} {(#2)} {#3}}}
\nc{\pra}[3]{{\it  Phys.\ Rev.\ }{{\bf A#1} {(#2)} {#3}}}
\nc{\prb}[3]{{\it  Phys.\ Rev.\ }{{\bf B#1} {(#2)} {#3}}}
\nc{\prc}[3]{{\it  Phys.\ Rev.\ }{{\bf C#1} {(#2)} {#3}}}
\nc{\prd}[3]{{\it  Phys.\ Rev.\ }{{\bf D#1} {(#2)} {#3}}}
\nc{\prl}[3]{{\it Phys.\ Rev.\ Lett.\ }{{\bf #1} {(#2)} {#3}}}
\nc{\pl}[3]{{\it  Phys.\ Lett.\ }{{\bf #1} {(#2)} {#3}}}
\nc{\prep}[3]{{\it Phys\. Rep.\ }{{\bf #1} {(#2)} {#3}}}
\nc{\prsl}[3]{{\it Proc.\ R.\ Soc.\ London\ }{{\bf #1} {(#2)} {#3}}}
\nc{\ptp}[3]{{\it  Prog.\ Theor.\ Phys.\ }{{\bf #1} {(#2)} {#3}}}
\nc{\ptps}[3]{{\it  Prog\ Theor.\ Phys.\ suppl.\ }{{\bf #1} {(#2)} {#3}}}
\nc{\physa}[3]{{\it  Physica\ A\ }{{\bf #1} {(#2)} {#3}}}
\nc{\physb}[3]{{\it  Physica\ B\ }{{\bf #1} {(#2)} {#3}}}
\nc{\phys}[3]{{\it Physica\ }{{\bf #1} {(#2)} {#3}}}
\nc{\rmp}[3]{{\it  Rev.\ Mod.\ Phys.\ }{{\bf #1} {(#2)} {#3}}}
\nc{\rpp}[3]{{\it Rep.\ Prog.\ Phys.\ }{{\bf #1} {(#2)} {#3}}}
\nc{\sjnp}[3]{{\it Sov.\ J.\ Nucl.\ Phys.\ }{{\bf #1} {(#2)} {#3}}}
\nc{\spjetp}[3]{{\it Sov.\ Phys.\ JETP\ }{{\bf #1} {(#2)} {#3}}}
\nc{\yf}[3]{{\it Yad.\ Fiz.\ }{{\bf #1} {(#2)} {#3}}}
\nc{\zetp}[3]{{\it Zh.\ Eksp.\ Teor.\ Fiz.\ }{{\bf #1} {(#2)} {#3}}}
\nc{\zp}[3]{{\it Z.\ Phys.\ }{{\bf #1} {(#2)} {#3}}}
\nc{\zpc}[3]{{\it Z.\ Phys.\ C\ }{{\bf #1} {(#2)} {#3}}}
\nc{\ibid}[3]{{\sl ibid.\ }{{\bf #1} {#2} {#3}}}

\def\vP{{\bf P}}
\def\vQ{{\bf Q}}
\def\vB{{\bf B}}
\def\vE{{\bf E}}

\def\vr{{\bf r}}
\def\vk{{\bf k}}
\def\vp{{\bf p}}
\def\vq{{\bf q}}

\def\vX{{\bf X}}
\def\vY{{\bf Y}}
\def\vSig{\hbox{\boldmath$\Sigma$}}
\def\valpha{\hbox{\boldmath$\alpha$}}

\def\Psibar{\overline{\Psi}}

\begin{document}
%

\thispagestyle{empty}
\hbox to\hsize{February 1997\hfil SUITP-96-17}
\hbox to\hsize{\hfil MPI-PTh/97-11}

\vspace*{8mm}

\begin{center}
{\Large\bf Neutrinos with Magnetic Moment:\\[2mm]
Depolarization Rate in Plasma}
\end{center}

\vspace*{5mm}

\bc
{\large
Per Elmfors$^{\rm a}$, Kari Enqvist$^{\rm b}$,
Georg Raffelt$^{\rm c}$ and G\"unter Sigl$^{\rm c,d}$}
\\[2mm]

\sl

$^{\sl a}$Fysikum, Box 6730, S-113 85 Stockholm, Sweden\\
        elmfors@physto.se\\
$^{\sl b}$Physics Department, P.O. Box 9, FIN-00014 University of 
Helsinki, Finland\\
        enqvist@pcu.helsinki.fi\\
$^{\sl c}$Max-Planck-Institut f\"ur Physik, F\"ohringer Ring 6, 
D-80805 M\"unchen, Germany\\
        raffelt@mppmu.mpg.de\\
$^{\sl d}$Department of Astronomy and Astrophysics, The University 
of Chicago\\ Chicago, Illinois 60637-1433, U.S.A.\\
        sigl@mppmu.mpg.de
\ec

\vfil
\bc
{\bf Abstract} \\
\ec
{\small
\begin{quotation}
\noindent
Neutrinos with a magnetic moment $\mu$ change their helicity when
interacting with an electromagnetic field.  Various
aspects of this effect have been described as spin precession,
spin-flip scattering, and magnetic Cherenkov radiation.  These
perspectives are unified in an expression for the $\nu_L\to\nu_R$
transition rate which involves the correlators of the electromagnetic
field distribution.  Our general formula corrects a previous result
and generalizes it to the case where the fields cannot be viewed as
classical and where the momentum transfers need not be small.  We
evaluate our result explicitly for a relativistic QED plasma and determine
the depolarization rate to leading order in the fine structure constant. 
Assuming
that big-bang nucleosynthesis constraints do not allow a right-handed
neutrino in equilibrium we derive the limit $\mu<6.2\EE{-11}\mu_B$ 
on the neutrino magnetic moment. 
Bounds on $\mu$ from a possible large scale magnetic fields are found to be 
more stringent even for very weak fields.
\end{quotation}}
\vfill

\eject

\normalsize
\setcounter{page}{1}

\Section{s:intro}{Introduction}

A neutrino or other neutral particle with a magnetic moment $\mu$ 
gets depolarized when traversing a random distribution of
electromagnetic fields as, for example, in a plasma of charged
particles. This effect could be important in stars or in the early
universe where the standard weak interactions produce only left-handed
neutrinos.  More than twenty years ago the depolarization effect was
considered as a possibility to solve the solar neutrino problem, but
for plausible values of $\mu$ the rate was found to be too
small~\cite{Radomski}.  In supernova cores, the left-handed neutrinos
are trapped while the helicity-flipped states could escape more easily
because they scatter only by the assumed magnetic-dipole
interaction. The depolarization effect would thus disturb the standard
picture of supernova theory so that the known properties of supernovae
and the observed SN~1987A neutrino signal allow one to derive certain
limits on Dirac magnetic moments~\cite{Supernova,Mohanty}. In the
early universe, the radiation density and thus expansion rate would
increase due to the new thermally excited degree of freedom, modifying
the predicted primordial light-element abundances. A comparison with
observations again allows one to derive limits
on~$\mu$~\cite{Morgan,FY,LoebS89}.

In most of these studies the depolarization rate was calculated from
the spin-flip scattering process $\nu_L+X\to X+\nu_R$ where $X$
represents electrons, positrons, or other charged particles and the
interaction is due to the assumed neutrino magnetic moment. In
addition there may be electromagnetic modes with a ``space-like''
dispersion relation which allow for the Cherenkov emission
$\nu_L\to\nu_R+\gamma$ and absorption $\gamma+\nu_L\to\nu_R$ and thus
contribute to the depolarization rate \cite{Radomski,Mohanty}.  This
is the case in supernova cores where the photon dispersion relation is
dominated by the nucleon magnetic moments or in nonrelativistic
plasmas, but it is not the case in the relativistic $e^+e^-$ plasma of
the early universe.  The rate for higher-order scattering processes
such as magnetic Compton scattering $\gamma+\nu\to\nu+\gamma$ is
proportional to $\mu^4$ and is thus ignored in the present discussion
which is limited to $\mu^2$ effects.  To order $\mu^2$ right-handed
neutrinos can also be produced by pair processes of the type
$\gamma\to\nu_L\bar\nu_R$ (plasmon decay) or $e^+ e^-\to
\nu_L\bar\nu_R$ (pair annihilation). In supernova cores or the early
universe with a large population of left-handed neutrinos the
$\nu_L\to\nu_R$ reactions always seem to be more important than the
pair processes which, however, dominate in normal stars which do not
have a population of trapped left-handed neutrinos.

Loeb and Stodolsky \cite{LoebS89} were the first to look at the
depolarization effect from a more classical perspective. They argued
that in a macroscopic magnetic field, left-handed neutrinos simply
spin-precess and that even in a microscopic distribution of random
fields the depolarization rate should be calculable as a suitable
ensemble average over the spin-precession formula. They found a result
which represents the depolarization rate in terms of certain
correlation functions of the electromagnetic fields.  In principle,
this correlator approach incorporates all electromagnetic effects to
order $\mu^2$ which lead to depolarization such as spin-flip
scattering and the Cherenkov processes.

These different approaches have co-existed in the literature with no
apparent attempt to compare them, to understand their relationship, or
to check their mutual consistency.  One problem is that Loeb and
Stodolsky's approach is entirely classical so that it is not obvious
if and how their formula is limited when applied to a plasma which
involves nonclassical (quantum) excitations of the electromagnetic
field.

We thus reconsider neutrino depolarization both classically in the
spirit of Loeb and Stodolsky (Sec.~\ref{s:classical}) and from a
quantum-kinetic perspective (Sec.~\ref{s:quantum}).  Put another way,
we derive the depolarization rate in terms of correlation functions of
the electromagnetic fields which may be quasi-classical or true
quantum variables.  Even in the classical limit our general result
involves terms which are absent in Loeb and Stodolsky's formula.

We also compare with the imaginary part of the neutrino self-energy
(Sec.~\ref{s:Im}) which has a simple relation to the depolarization
rate. This calculation amounts to a determination of the dominant
spin-flip scattering rate, but even to lowest order in the
fine-structure constant $\alpha$ the correct screening prescription
has to be incorporated by resumming the photon propagator.  Then it
automatically includes all order $\mu^2$ contributions to the
production rate of right-handed neutrinos, i.e.\ it includes
$\nu_L\to\nu_R$ transitions from spin-flip scattering or from
Cherenkov processes as well as plasmon decay and $e^+e^-$
annihilation.
\bort{
However, in the relativistic plasma 
we shall only deal explicitly with the dominant
processes which do not include $e^+e^-$ annihilation.
It does not include pair processes which are higher
order in $\alpha$ such as the photo-neutrino process $\gamma+e^-\to
e^-+\nu_L+\bar\nu_R$.
}

In Sec.~\ref{s:plasma} we evaluate the depolarization rate explicitly
by using the correlators for a relativistic QED plasma and compare the
classical and quantum treatments.  In Sec.~\ref{s:bigbang} we apply
this result to the depolarization of neutrinos in the early universe,
allowing us to derive a limit on the magnetic dipole moment from
considerations of big-bang nucleosynthesis.  We also compare this
limit to previous bounds obtained from assuming a large-scale
background magnetic field. In Sec.~\ref{s:summary} we summarize our
findings.

\Section{s:classical}{Classical Trajectory Approach}

\subsection{Depolarization Rate}

A nonrelativistic neutral particle with a magnetic moment $\mu$ and an
intrinsic angular momentum (spin) precesses in the presence of an
external magnetic field $\vB$ according to
$\dot\vP=g\mu\,\vB\times\vP$. Here, $\vP$ is the spin polarization
vector which, for a spin-$\inv{2}$ particle, 
parametrizes the spin density matrix in the usual form
$\rho=\inv{2}(1+P_i\sigma_i)$.  Further, $g$ is the gyromagnetic
ratio.  Loeb and Stodolsky~\cite{LoebS89} used $g=1$, i.e.\ a
classical particle, while we always study neutrinos with spin
$\frac{1}{2}$ so that $g=2$. If the particle moves relativistically,
the main modification is that in the laboratory frame only the
magnetic field transverse to the direction of motion contributes, and
that a transverse electric field is also important because in the
particle's rest frame it manifests itself as a magnetic
field. Altogether, the precession formula for an ultrarelativistic
($v=1$) spin-$\frac{1}{2}$ particle is 
\be{precession}
        \dot\vP=2\mu(\vB_\perp-\hat\vp\times\vE_\perp) \times\vP ~~,
\ee 
where the subscript $\perp$ refers to the field vectors transverse to
the direction of motion and $\hat\vp$ is a unit vector in the
direction of the neutrino momentum.

The same equation of motion is obtained if one begins directly with
the covariant Lagrangian which describes the coupling of a Dirac
neutrino with magnetic moment $\mu$ to the electromagnetic field
tensor $F^{\mu\nu}$,
\be{L} 
\cL_{\rm int}=-{\textstyle{1\over2}}\,\mu
\,\Psibar\sigma_{\mu\nu}F^{\mu\nu}\Psi
=\mu\Psibar\,(\vSig\cdot\vB-i\valpha\cdot\vE)\,\Psi~~, 
\ee 
where $\Psi$ is the neutrino Dirac field.  In terms of the usual Dirac
matrices we have $\Sigma_i=\gamma_5\gamma^0\gamma_i$ and
$\alpha_i=\gamma^0\gamma_i$.  This Lagrangian shows that the
magnetic-moment interaction indeed only couples neutrinos of opposite
chirality. As we are  concerned only with ultrarelativistic
neutrinos this is tantamount to a coupling of opposite helicities.

In the presence of other forces than those represented by the
electromagnetic fields, the off-diagonal elements of the density matrix 
can be damped by
collisions which ``measure'' the helicity content of a neutrino
state. Standard weak interactions would have this property because
only left-handed neutrinos are affected by collisions. This damping
effect is important when the spin precesses in macroscopic magnetic
fields~\cite{ElmforsGR96,EnqvistRS95}, and can be taken into account
by adding a term $-D\,P_i$ to the right hand side of
the equation of motion for $\dot P_i$, $i=1,2$, in 
\eq{precession}.  For depolarization in stochastic electromagnetic
fields one can speak of a coherent spin precession only for a time
period which represents the correlation time of the 
fields. The damping effect would be important if the average time
between weak collisions were shorter than this coherence time
scale. An example is a putative horizon-scale magnetic field in the
early universe.

We always assume ultrarelativistic (i.e.\ effectively massless)
neutrinos. However, in a medium the dispersion relations
for left- and right-handed states are different because only the
left-handed ones feel the weak potential produced by the background
medium.  We may write the refractive energy difference between left-
and right-handed states in the form $\omega_{\rm refr}=2\mu B_{\rm
refr}$ in terms of an effective magnetic field $B_{\rm refr}$ which
points in the neutrino's direction of motion. Therefore, the
precession formula $\dot\vP=2\mu\vB^{\rm eff}\times\vP$ finally
involves an effective magnetic field with $\vB^{\rm
eff}_\perp=\vB_\perp-\hat\vp\times\vE_\perp$ and $\vB^{\rm
eff}_\parallel=(\omega_{\rm refr}/2\mu)\hat\vp$.

In general, the electric and magnetic fields depend on location and
time in arbitrary ways.  Therefore, a neutrino with a magnetic moment
will be deflected. However, the deflecting forces on a magnetic dipole
are proportional to the field gradients while the precession effect
depends on the fields directly. Therefore, if the spatial variations
are small the neutrinos can still
be assumed to move on a straight line which can be taken to be the
$z$-direction. Moreover, the neutrino can be taken to ``see'' only the
fields at a specific location which is assumed to vary with time as
$z=t$ because of the assumed propagation with the speed of light and
because we take $z=0$ at $t=0$. Therefore, through the condition $z=t$
the fields $\vB^{\rm eff}$ are to be viewed as functions of time
alone.  We call this the ``classical trajectory approach'' to the
problem of neutrino spin depolarization.

In order to derive the equation of motion of the polarization vector
in a stochastic field distribution it proves useful to write the
precession equation in the form 
\be{depeq}
   \dot\vP(t)=M(t)\,\vP(t)~~,
\ee
where the time-dependent matrix $M$ is explicitly given by
\be{M} M_{ij}=
\left(\ba{ccc}-D&-\omega_{\rm refr}&2\mu(B_y-E_x)\\ 
              \omega_{\rm refr}& -D & -2\mu(B_x+E_y)\\ 
             -2\mu(B_y-E_x)&2\mu(B_x+E_y) &0 \ea\right) 
\ee
if the neutrinos are ultrarelativistic and move in the $z$-direction. 

In general we must view the matrix $M$ as consisting of one part 
$\<M\>$ which is nonzero on the average plus stochastic fluctuations
around that average. Even if there are no large-scale magnetic fields,
the refractive effect provides a nonvanishing average contribution. In
order to derive the depolarization rate we eliminate $\< M\>$ by going
to an ``interaction picture'' with $\vQ(t)\equiv e^{-\<M\>t}\,\vP(t)$
so that we are left with the equation of motion
\be{Qeq}
    \dot\vQ(t)=m(t)\vQ(t)
\ee
with
\be{mdef}
    m(t)\equiv e^{-\<M\>t}(M-\<M\>)e^{\<M\>t}~~.
\ee
The formal solution to \eq{Qeq} is
\be{solQeq}
        \vQ(t)=\sum_{n=0}^\infty \int_0^tdt_1\cdots
        \int_0^{t_{n-1}}dt_n\,
        m(t_1)\cdots m(t_n)\, \vQ(0)~~.
\ee
This is the solution we need to average over a statistical ensemble of
field configurations.
We shall assume that the field fluctuations are Gaussian so that the
$n$-point correlation functions can be reduced to 
products of pair correlation functions.
Let us observe that in QED the high temperature effective action is
in fact Gaussian \cite{BraatenP90}. (This is not a general
feature and not true e.g. for QCD.) We seek the solution only for times $t$
much larger than the correlation time of the stochastic fields.  
In the integral in \eq{solQeq} all time arguments have to occur in close
pairs for the integrand to be non-negligible.  Since the time
arguments are ordered it is only the permutation where adjacent
matrices are contracted that gives any contribution in the leading
large-time limit.  It can also be checked at the end that this
asymptotic region is reached long before one decay time in the present
case of neutrino spin oscillations.
We can therefore approximate $\<\vQ(t)\>$ by
\be{Qave}
        \<\vQ(t)\>\approx \sum_{n=0}^\infty \int_0^tdt_1\cdots
           \int_0^{t_{n-1}}dt_n\,
        \<m(t_1)m(t_2)\>\cdots \<m(t_{n-1})m(t_n)\>\,\vQ(0)~~.
\ee
After computing the pair correlation matrix $\<m(t_1)m(t_2)\>_{ij}$
with zero average field strength 
we find that there is no mixing between the third component and the 
rest so that it is useful to define
\bea{ndef}
        n(t_1-t_2)&\equiv&-\<m(t_1)m(t_2)\>_{33}\\
        &=&(2\mu)^2e^{-D(t_1-t_2)}\biggl\{
        \cos[\omega_{\rm refr}(t_1-t_2)]
        \Bigl\langle
        \vB_\perp(t_1)\cdot\vB_\perp(t_2)+
        \vE_\perp(t_1)\cdot\vE_\perp(t_2)\nonumber\\
        &&\hskip14em
        +\,\hat\vp\cdot
        \Bigl[\vB(t_1)\times\vE(t_2)-\vE(t_1)\times\vB(t_2)\Bigr]
        \Bigr\rangle\nonumber\\
        &&\hskip6em
        +\,\sin[\omega_{\rm refr}(t_1-t_2)]
        \Bigl\langle
        \vE_\perp(t_1)\cdot\vB_\perp(t_2)-
        \vB_\perp(t_1)\cdot\vE_\perp(t_2)\nonumber\\
        &&\hskip14em
        +\,\hat\vp\cdot
        \Bigl[\vB(t_1)\times\vB(t_2)-\vE(t_1)\times\vE(t_2)\Bigr]
        \Bigr\rangle\biggr\}~~.\nonumber
\eea
Of course, in the terms involving cross products of fields we could
have equally used the transverse field components. 

The depolarization rate is given by the shrinking rate of
$\<P_3(t)\>$.
If we carry out all the integrals in \eq{Qave}, keeping only the
leading term at large $t$, we finally find that the ensemble-averaged
evolution of $P_3$ is given by
\be{pevolution}
        \<P_3(t)\>=\exp\left[-t\int_0^\infty dt'\,n(t')\right]P_3(0)
        \equiv e^{-\Gamma_{\rm depol} t}P_3(0)~~.
\ee

If $\omega_{\rm refr}$ is much larger than the inverse correlation
time of the electromagnetic fields, the oscillating $\cos(\omega_{\rm
refr} t)$ and $\sin(\omega_{\rm refr} t)$ terms under the integral in
\eq{ndef} would suppress the depolarization effect. Likewise, a very
large helicity-measuring damping coefficient $D$ would prevent
oscillations and suppress the depolarization.  In stars, even
supernovae, and in the early universe these suppression effects are
insignificant because the inverse of the electromagnetic correlation
length is much larger than both the refractive energy difference
between those states and the damping rate.  Therefore, for all
situations which are of astrophysical interest we can put $\omega_{\rm
refr}=D=0$.  (In the presence of a large-scale background field there
is an additional component to $\Gamma_{\rm depol}$ from $\<M\>$ 
causing a coherent
spin precession. Of course, this additional term would depend on both
$\omega_{\rm refr}$ and $D$; see Sec.~\ref{ss:lsB}.) 
The depolarization rate from stochastic fields is then found to be
\bea{depolrate}
 \Gamma_{\rm depol}&=&(2\mu)^2\int_0^\infty dt\,
\Bigl\langle\vB_\perp(t)\cdot\vB_\perp(0)
+\vE_\perp(t)\cdot\vE_\perp(0)\nonumber\\
&&\hskip6em
+\,\hat\vp\cdot
\Bigl[\vB(t)\times\vE(0)-\vE(t)\times\vB(0)\Bigr]\Bigr\rangle~~.
\eea
This result agrees with that of Loeb and Stodolsky \cite{LoebS89}
except for the cross term.  They agree that it should be there and
stress that it can be derived rather easily by beginning with the
spin-precession formula in the neutrino's rest frame where only the
$\vB_\perp^2$ correlator appears, and express it by the Lorentz
transformed laboratory fields.%
\footnote{L.~Stodolsky, private communication.}

Two remarks about the interpretation of \eq{depolrate} are in order.
First, a term like $\vB(t)$ really means $\vB(t,\vr)$ with
$\vr=\hat\vp t$ where $\hat\vp$ is the neutrino velocity
vector. Therefore, even in an isotropic medium a vectorial quantity
like $\<\vE(t)\times\vB(0)\>$ need not vanish because it depends on
the external vector $\hat\vp$.  Second, the integrand is an even
function of $t$ because the fields are classical variables so that
their ordering is arbitrary, and because of time translational invariance
of the statistical ensemble in equilibrium. 
Therefore, the time integral may be
extended to $-\infty$ if a compensating factor $1/2$ is introduced.
We will always use this more symmetric form which is easier to
handle in Fourier space, and which allows for a direct transition to
the quantum case (Sec.~\ref{s:quantum}) where the ordering of the
fields is important.

\Subsection{ss:iso}{Isotropic Medium}

In a homogeneous and isotropic 
plasma or other medium, correlator expressions like the ones
appearing in \eq{depolrate} are conveniently calculated in Fourier
space. To this end we introduce the usual notation\footnote{We 
use $K=(k_0,\vk)$, $P=(p_0,\vp)$ etc.\ for four-vectors and 
$k=|\vk|$, $p=|\vp|$ etc.\ for the corresponding three-vectors.}  
\be{corrdef}
\<X_iY_j\>_K\equiv
\int_{-\infty}^{+\infty}dt\,e^{ik_0 t}\<X_i(t,\vk) Y_j(0,-\vk)\>
\ee
where $X_i$ and $Y_j$ each represent a component of the $E$- or 
$B$-fields with $X_i(t,\vk)$ the spatial Fourier transform of 
$X_i(t,\vr)$ and so forth.
Therefore, the depolarization rate \eq{depolrate} can be expressed as
\be{depolrate2}
\Gamma_{\rm depol}=\frac{(2\mu)^2}{2}
\int_{-\infty}^{+\infty} dt
\int\frac{d^3\vk}{(2\pi)^3}\int_{-\infty}^{+\infty}
\frac{dk_0}{2\pi}\,
e^{i(\hat\vp\cdot\vk-k_0t)}\,S_P(K)~~,
\ee
where the $P$-dependent ``dynamical structure function'' is 
\be{Sdef}
S_P(K)\equiv
\Bigl\<\vB_\perp^2+\vE_\perp^2
+\hat\vp\cdot(\vB\times\vE-\vE\times\vB)\Bigr\>_K~~.
\ee
It will turn out that $\mu^2 S_P(K)$ is to be interpreted as
the probability for a neutrino of four 
momentum $P$ to transfer the four momentum
$K$ to the medium. 
The $dt$ integration yields
$2\pi\delta(\hat\vp\cdot\vk-k_0)$ which allows us to perform the
$dk_0$ integration trivially. Altogether we find
\be{depolrate3}
\Gamma_{\rm depol}=2\mu^2
\int\frac{d^3\vk}{(2\pi)^3}\,S_P(K)~~,
\ee
where $K$ is restricted by the condition $k_0=\hat\vp\cdot\vk$.

The structure of $S_P$ still depends on the field components 
transverse
to the neutrino momentum. We stress that contrary to
Ref.~\cite{LoebS89} the assumed isotropy of the medium does not imply
that $\<\vB_\perp^2\>_K$ equals $\frac{2}{3}\<\vB^2\>_K$ because
$\<\vB_\perp^2\>_K$ depends on the external vector $\hat\vp$
and thus is not a scalar. In order to use isotropy
properly we observe that $\vB_T=\vB-(\hat\vp\cdot\vB)\hat\vp$ so that
\be{Stensor}
S_P(K)=
\<B_iB_j+E_iE_j\>_K\,(\delta_{ij}-\hat p_i\hat p_j)
+\<B_iE_j-E_iB_j\>_K\,\epsilon_{ij\ell}\hat p_\ell~~.
\ee
We further note that in an isotropic medium any correlator expression
of the form $\<X_iY_j\>_K$ can be proportional only to
$\delta_{ij}$, $\hat k_i\hat k_j$, or $\epsilon_{ij\ell}\hat k_\ell$
because only the vector $\hat\vk$ is available to construct spatial
tensor structures. The most general structure is thus found to~be
\bea{xystruc}
\<X_iY_j\>_K&=&
\<\vX\cdot\vY\>_K\,\frac{\delta_{ij}-\hat k_i\hat k_j}{2}
+\<(\vX\cdot\hat\vk)(\vY\cdot\hat\vk)\>_K\,
\frac{3\hat k_i\hat k_j-\delta_{ij}}{2}\nonumber\\
&&+\,\<\vX\times\vY\>_K\cdot\hat\vk\,\,
\frac{\epsilon_{ij\ell}\hat k_\ell}{2}~~.
\eea
It is then straightforward to show that
\bea{Stensor2}
S_P(K)&=&\<\vB^2+\vE^2\>_K
\frac{1+(\hat\vk\cdot\hat\vp)^2}{2}
+\<(\vB\cdot\hat\vk)^2+(\vE\cdot\hat\vk)^2\>_K\,
\frac{1-3(\hat\vk\cdot\hat\vp)^2}{2}\nonumber\\
&&+\,(\hat\vk\cdot\hat\vp)\,
\<\vB\times\vE-\vE\times\vB\>_K\cdot\hat\vk~~.
\eea
This expression simplifies further if we observe that 
$\vE$ and $\vB$ must 
obey Maxwell's equations. Because the magnetic field is divergence 
free we have $\vB\cdot\vk=0$. Further, $\<(\vE\cdot\vk)^2\>_K=
\<\vE^2\vk^2-(\vk\times\vE)^2\>_K$. Maxwell's equations give us
$\vk\times\vE=k_0\vB$ so that 
$\<(\vE\cdot\hat\vk)^2\>_K=\<\vE^2\>_K-(k_0/k)^2\<\vB^2\>_K$. 
Next, $\<\vB\times\vE\>_K\cdot\vk=\<\vB\cdot(\vE\times\vk)\>_K$, 
allowing us again to apply $\vk\times\vE=k_0\vB$ so that
$\<\vB\times\vE-\vE\times\vB\>_K\cdot\hat\vk=
-2(k_0/k)\,\<\vB^2\>_K$.
Altogether we thus find
\be{Stensor3}
S_P(K)=\<\vB^2\>_K
\left(\frac{1+(\hat\vp\cdot\hat\vk)^2}{2}-\frac{2k_0}{k}\,
\hat\vp\cdot\hat\vk
-\frac{k_0^2}{k^2}\frac{1-3(\hat\vp\cdot\hat\vk)^2}{2}\right)
+\<\vE^2\>_K \left[1-(\hat\vp\cdot\hat\vk)^2\right]~~.
\ee
In an isotropic medium the $\<\ldots\>_K$ expressions
depend only on $(k_0,k)$ and not on the direction $\hat\vk$.

In order to present our final result we note that the depolarization
rate studied thus far represents the rate by which a fixed
ensemble of neutrinos gets depolarized. In the early universe or in
supernovae, however, a more relevant quantity is the spin-flip rate
which measures the speed by which the sea of right-handed neutrinos
gets populated if the number density of left-handed neutrinos is held
fixed at its thermal equilibrium value because they are replenished
by other reactions. Evidently, since the number of right- and left-handed
neutrinos is $n_{R,L}=\inv{2}(1\pm P_3)$ we have that  $\dot
n_R/n_L\equiv\Gamma_{\rm flip}=\Gamma_{\rm depol}/2$ so that 
\be{Gflip}
\Gamma_{\rm flip}=\mu^2
\int\frac{d^3\vk}{(2\pi)^3} 
\left[1-(\hat\vp\cdot\hat\vk)^2\right]
\left(\<\vE^2\>_K+\frac{1-3(\hat\vp\cdot\hat\vk)^2}{2}
\<\vB^2\>_K\right)~~,
\ee
where, again, $K$ is constrained by $k_0=\hat\vp\cdot\vk$.

\Section{s:quantum}{Quantum Kinetic Approach}

\subsection{Relaxation Rate}
\label{ss:RelaxationRate}

The spin-flip rate derived by the classical-trajectory approach in the
previous section is valid only in the approximation that the momentum
transfer $\vk$ is always much smaller than the neutrino momentum $\vp$
so that it is justified to view the neutrino as propagating on a
straight line during a typical field correlation time. Further, the
electric and magnetic fields were taken to be classical variables,
ignoring the quantized nature of the exchange of energy and
momentum $(k_0,\vk)$ between the neutrino and the medium.

In a quantum-kinetic approach the relaxation of the
neutrino helicity by electromagnetic interactions is described by a
Boltzmann collision equation of the form
\be{BCE}
\partial_t f_{\vp}^R=\int \frac{d^3\vq}{(2\pi)^3}
\left[W_{L\to R}(Q,P) f_{\vq}^L (1-f_{\vp}^R)-
W_{R\to L}(P,Q) f_{\vp}^R (1-f_{\vq}^L)+\ldots\,\right]
\ee
where $f^{L,R}_{\vp}$ are the occupation numbers for left- and
right-handed neutrinos of momentum $\vp$, respectively.  Here,
$W_{R\to L}(P,Q)$ is the transition rate for a right-handed neutrino of
four momentum $P$ to a left-handed one of $Q$ under the influence of
the electromagnetic fields of the medium. There are also terms
representing pair production and annihilation processes which we will
discuss below.  If there are large-scale magnetic fields the Boltzmann
equation includes an oscillation term analogous to the simultaneous
treatment of collisions and flavor oscillations in Refs.~\cite{RSS}.

The neutrino interaction with the electromagnetic fields is of the
current-current form. It is well known from linear-response theory
that in this situation the transition probability is given essentially
by the dynamical structure function of the medium (in our case the
electromagnetic fields), i.e.\ by correlator expressions like the
classical ones discussed in the previous section. Therefore, all that
remains to be done to derive $W(P,Q)$ is to determine the required
tensorial contraction between the neutrino momenta and the
electromagnetic field correlators.   

To this end we begin with the interaction Lagrangian
\eq{L} and consider the matrix element for the transition of a
left-handed neutrino with four momentum $P$ to a right-handed one
with $Q$. With the massless neutrino Dirac spinors 
$u_{P,L}$ and $u_{Q,R}$ which involve the appropriate chirality
projections one finds 
\hbox{${\cal M}=\mu\,\overline u_{P,L}(\vB\cdot\vSig-i\vE\cdot\valpha)
u_{Q,R}$}. Of course, because the interaction couples only states of
opposite chirality it would have been enough to include one chirality
projector. Taking the square of this matrix element and carrying out
the Dirac traces we find  
\bea{WLR}
\mu^{-2}W_{L\to R}(P,Q)&=&
\<B_iB_j+E_iE_j\>_K\,
\frac{(1+\hat\vp\cdot\hat\vq)\delta_{ij}-\hat p_i\hat q_j
-\hat q_i\hat p_j}{2}\nn
&&+\,\<B_iE_j-E_iB_j\>_K\,\epsilon_{ij\ell}\,
\frac{\hat p_\ell+\hat q_\ell}{2}\nn
&&+\,i\<E_iE_j\>_K\,\epsilon_{ij\ell}\,
\frac{\hat p_\ell+\hat q_\ell}{2}~~,
\eea
where $K=P-Q$.
In the limit of small $\vk$ we may set $\hat\vp=\hat\vq$, taking us
back to \eq{Stensor} of the classical-trajectory approach apart from
the new term which involves $\<\vE\times\vE\>_K$. It is easy to show
that in the classical limit where the fields are commuting variables
this term disappears under the $d^3\vk$ phase-space integration so
that our present result leads to the same classical spin-flip rate.

In the general case the status of the $\vE\times\vE$ term is more
subtle. It has the opposite sign for an $R\to L$ transition. On the
other hand it is easy to show that $W_{L\to R}(P,Q)=W_{R\to L}(P,Q)$
if the medium is invariant under parity. Therefore, in such a medium
the $\<\vE\times\vE\>_K$ term must vanish even in the general
case. This is not surprising in the sense that the electromagnetic
properties of a parity invariant medium are characterized by precisely
two independent ``material constants'' which can be chosen to be the
dielectric permittivity and the magnetic permeability. Another pair of
equivalent parameters are the longitudinal and transverse polarization
functions $\Pi_{T,L}(K)$ which we will use below. Therefore, there are
only two independent field correlator expressions, e.g.\ $\<\vE^2\>_K$
and $\<\vB^2\>_K$. They are related to 
$\Pi_{T,L}(K)$ by the fluctuation-dissipation theorem.  A
parity-noninvariant medium, on the other hand, is characterized by
three independent parameters---there are two different transverse
polarization functions. The third function gives rise to a
nonvanishing $\<\vE\times\vE\>_K$ field correlator.

For the neutrino spin relaxation problem we inevitably have a medium
which is not parity invariant because initially it involves only
left-handed neutrinos. By their assumed magnetic moments they are
expected to give rise to a nonvanishing $\<\vE\times\vE\>$ term, i.e.\
$L\to R$ and $R\to L$ collisions are not expected to occur with the
same transition probability. However, in terms of the relaxation rate
this would be an order $\mu^4$ effect so that to order $\mu^2$ we may
ignore electromagnetic neutrino-neutrino interactions. In the cases of
interest to us we may thus ignore the $\<\vE\times\vE\>_K$ term and
use $W(P,Q)\equiv W_{L\to R}(P,Q)=W_{R\to L}(P,Q)$ for the transition
rate.

We proceed by applying Maxwell's equations in the same way as in the
previous section, never changing the order in which the non-commuting
field operators appear. Since we never changed this order even
in the classical case, we find the same simplifications as there. The
final contraction of indices then yields
\bea{Wquantum}
\mu^{-2}W(P,Q)&\!=\!&
\<\vB^2\>_K\!\left(
\frac{1+(\hat\vp\cdot\hat\vk)(\hat\vq\cdot\hat\vk)}{2}
-\frac{2k_0}{k}\frac{\hat\vp\cdot\hat\vk+\hat\vq\cdot\hat\vk}{2}
-\frac{k_0^2}{k^2}
\frac{\hat\vp\cdot\hat\vq-3(\hat\vp\cdot\hat\vk)(\hat\vq\cdot\hat\vk)}{2}
\right)\nn
&&+\,\<\vE^2\>_K\left(
\frac{1+\hat\vp\cdot\hat\vq}{2}
-(\hat\vp\cdot\hat\vk)(\hat\vq\cdot\hat\vk)\right)~~.
\eea
This complicated looking expression can be transformed
to
\be{Wquantum2}
W(P,Q)=-\mu^2\,
\frac{K^2(p_0+q_0)^2}{8k^2 p_0 q_0}
\left[2\<\vE^2\>_K+ 
\<\vB^2\>_K\left(1-\frac{3k_0^2}{k^2}
+\frac{K^2}{(p_0+q_0)^2}
\right)\right]~~.
\ee
Note that $q_0=p_0-k_0$ and $p_0=p$ and that $-K^2=-(P-Q)^2
=2PQ=2(p_0q_0-\vp\cdot\vq)$ is a positive number.

We now return to the Boltzmann collision equation~(\ref{BCE}) for a
parity invariant medium where we need only one function $W(P,Q)$. The 
transition rate for the pair production and annihilation processes is
given by the same function with ``crossed'' four momenta, i.e.\ 
$P\to -P$ or $Q\to-Q$. The complete collision equation is then
\bea{BCEfull}
\partial_t f_{\vp}^R&=&\int \frac{d^3\vq}{(2\pi)^3}\,
\Bigl[W(Q,P) f_{\vq}^L (1-f_{\vp}^R)-
W(P,Q) f_{\vp}^R (1-f_{\vq}^L)\nn
&&\hskip5em+\,W(-Q,P)(1-f_{\vq}^L)(1-f_{\vp}^R)
-W(Q,-P)f_{\vq}^L f_{\vp}^R\Bigr]~~.
\eea  
Here, the first term in the collision integral represents gain by
$L\to R$ spin-flip scattering, the second loss by $R\to L$ scattering,
the third gain by pair production, and the fourth loss by pair
annihilation. 

In the astrophysical applications that we are interested in
(supernovae, early universe) the left-handed neutrinos are and remain
in thermal equilibrium so that we may replace
$f_\vp^L$ by the Fermi-Dirac distribution $f_\vp^F$
at the appropriate temperature and chemical potential. The collision
equation is then a linear differential equation of the form
\be{BCEgainloss}
\partial_t f_{\vp}^R=
\Gamma_{\rm gain}(\vp)(1-f_{\vp}^R)-\Gamma_{\rm loss}(\vp)f_{\vp}^R 
\ee 
with
\bea{gainloss}
\Gamma_{\rm gain}(\vp)&=&
\int \frac{d^3\vq}{(2\pi)^3}
\left[W(Q,P) f_{\vq}^F+W(-Q,P)(1-f_{\vq}^F)\right]~~,\nn
\Gamma_{\rm loss}(\vp)&=&
\int \frac{d^3\vq}{(2\pi)^3}
\left[W(P,Q) (1-f_{\vq}^F)+W(Q,-P)f_{\vq}^F\right]~~.
\eea 
Further, in equilibrium we have $\partial_t f_{\vp}^R=0$ and
$f_{\vp}^R=f_\vp^F$. Inserting this in \eq{BCEgainloss} reveals that
$\Gamma_{\rm gain}=\Gamma_{\rm tot} f_\vp^F$ 
with $\Gamma_{\rm tot}=\Gamma_{\rm gain}+\Gamma_{\rm loss}$
so that we may write
\be{BCEgainlossfinal}
\partial_t (f_{\vp}^R-f_\vp^F) =
-\Gamma_{\rm tot}(\vp)(f_{\vp}^R-f_\vp^F)~~. 
\ee 
Therefore, $\Gamma_{\rm tot}(\vp)$ is the appropriate rate
that measures the exponential approach of the $\vp$ mode to helicity
equilibrium.

\subsection{Correlation Functions}
\label{ss:corrfunc}

In order to evaluate the relaxation rate we need to know the electric
and magnetic field fluctuations in a given medium. 
By virtue of the fluctuation-dissipation theorem they are related to
the imaginary part of the medium's dielectric response functions. One
way of expressing these quantities is in terms of
the longitudinal and transverse photon spectral
functions $\cA_{T,L}$ which are
the coefficients in the decomposition
$\cA_{\mu\nu}=-\cP_{\mu\nu}\cA_T-\cQ_{\mu\nu}\cA_L$ in the Landau gauge.
The photon spectral function $\cA_{\mu\nu}(K)$ is
related to the retarded and advanced Green's functions through
$\cA_{\mu\nu}(K)=
[iD_{\mu\nu}(k_0+i\epsilon,\vk)
-iD_{\mu\nu}(k_0-i\epsilon,\vk)]/2\pi$.
The transverse and longitudinal projection operators 
have the nonvanishing components 
$\cP_{ij}(K)=-\delta_{ij}+(k_ik_j)/k^2$ and
$\cQ_{\mu\nu}(K)=-(K^2k^2)^{-1}
(k^2,-k_0\vk)_\mu (k^2,-k_0\vk)_\nu$. 
In terms of the photon polarization functions we have
\be{ATL}
   \cA_{T,L}(K)=-\inv{\pi}\,
        \frac{\Im \Pi_{T,L}}
        {|K^2-\Re \Pi_{T,L}|^2+|\Im \Pi_{T,L}|^2}~~.
\ee
The analytic continuation in the imaginary part is the retarded
$\Im\Pi_{T,L}(k_0+i\epsilon)$.
Note that the photon polarization tensor is given by 
$\Pi_{\mu\nu}=\cP_{\mu\nu}\Pi_T+\cQ_{\mu\nu}\Pi_L$.

In terms of the spectral functions the field fluctuations are found to
be~\cite{Lemoine95}
\bea{BBEE2}
\<\vB^2\>_K&=&\frac{2\pi}{1-e^{-\beta k_0}}
        2k^2\cA_T(K)~~,\nn
\<\vE^2\>_K&=&\frac{2\pi}{1-e^{-\beta k_0}}
        \left[2k_0^2\cA_T(K)+K^2\cA_L(K)\right]~~,
\eea
where $\beta=1/T$ is the inverse temperature. Note that for a positive
$k_0$ (energy given to the medium) the thermal factor is identical
with $1+(e^{\beta k_0}-1)^{-1}$, i.e.\ it is understood as a Bose
stimulation factor for exciting a quantum $(k_0,\vk)$ of the
medium. Conversely, if $k_0<0$ (energy lost by the medium) this factor
is $-(e^{\beta |k_0|}-1)^{-1}$. Apart from the minus sign it is the
occupation number for such an excitation. The functions $\cA_{T,L}(K)$
are odd in $k_0$ so that the negative sign of the thermal factor is
automatically absorbed. Put another way, the correlators obey
detailed-balance conditions of the form
 $\<\vB^2\>_{-K}=\<\vB^2\>_{K} e^{-\beta k_0}$.

\Section{s:Im}{Imaginary Part of the Neutrino Self-Energy}

An alternative way of calculating the rate of populating right-handed
neutrinos is through the imaginary part of the neutrino self-energy
\cite{Weldon83}. The relaxation rate $\Gamma_{\rm tot}$ introduced in
the previous section is directly related to the imaginary part of the
neutrino self-energy through
\be{GIms}
\Gamma_{\rm tot}(\vp)=-\frac{2\,\Im \overline{u}_{P,R}
\Sigma(p_0+i\epsilon) u_{P,R}}{\overline{u}_{P,R}u_{P,R}}~~,
\ee
where $u_{P,R}$ is the Dirac spinor of a right-handed neutrino with
four momentum $P$. 

The self-energy to one-loop order 
is a bubble diagram with a left-handed
neutrino and  a photon propagator connected via the 
interaction vertex $\mu\,K_\alpha\sigma^{\alpha\mu}$. Its time
ordered imaginary part can be shown to be \cite{KobesS85}
\bea{ImS}
        \Im \Sigma(P)&=&-\frac{\mu^2\sign(p_0)}{\sin 2\phi_P}
        \int \dbar{4}{K}\,\sign(p_0+k_0)\sign(k_0)
        \inv{2}\sin2\phi_{P+K}\inv{2}\sinh 2\theta_K \nn
        &&\times K_\alpha\sigma^{\alpha\mu}
        (\slask P+\slask K)\inv{2}(1-\g_5)K_\beta\sigma^{\beta\nu} 
        (2\pi)^2\delta((P+K)^2)\cA_{\mu\nu}(K)~~,
\eea
where $\epsilon(x)=\pm 1$ depending on the sign of $x$, 
the photon spectral function $\cA_{\mu\nu}$ was defined in
Sec.~\ref{ss:corrfunc} above, and  
\be{sindef}
   \inv{2}\sin 2\phi_K=\frac{e^{\beta|k_0|/2}}{e^{\beta|k_0|}+1}~~,
        \qquad
   \inv{2}\sinh 2\theta_K=\frac{e^{\beta|k_0|/2}}{e^{\beta|k_0|}-1}~~.
\ee

The neutrino inside the loop is necessarily left handed if the
external one is right handed since the interaction flips chirality, so
we do not need to include the $\inv{2}(1+\g_5)$ part of the propagator
to compute the right-handed self-energy.  The fermion distribution
function in \eq{sindef} should therefore describe a fully populated
equilibrium ensemble for which the chemical potential is taken to
vanish.
\bort{
In the evaluation of the neutrino matrix element we have used
spin-averaged spinors because in a parity conserving medium the
interaction rate of a left-handed or right-handed neutrino is the
same. Put another way, from the start we have neglected the
$\vE\times\vE$ contribution that would appear in a parity noninvariant
system---see the discussion after~\eq{WLR}.
}

The results of Refs.~\cite{Weldon83} and \cite{KobesS85} differ by
the overall factor $\sign(p_0)$ which is related to the retarded
\cite{Weldon83} or
time ordered \cite{KobesS85} prescription for $\Im \Sigma$. The
physical sign is however obvious since $\Gamma_{\rm tot}$ must be 
positive so we can concentrate on $p_0>0$.
Explicitly we find for  an on-shell $\nu_R$
with momentum $p$
\bea{GfIm}
        \Gamma_{\rm tot}(p)&=& \frac{\mu^2}{2\pi}\int_0^\infty dk\,k
        \int_{-\infty}^\infty dk_0 
        \,\theta\biggl(-K^2(K^2+4pk_0+4p^2)\biggr)\nn
        &&\times
        \left[\frac{\sign(k_0)}{e^{\beta|p_0+k_0|}+1}
        +\frac{\sign(p_0+k_0)}{e^{\beta|k_0|}-1}+\sign(p_0+k_0)
         \theta(-k_0)
        -\sign(k_0)\theta(-p_0-k_0)\right]\nn
        &&\times\frac{K^4}{k^2}
        \left[\left(1+\frac{k_0}{p}+\frac{K^2}{4p^2}\right)\cA_T(K)
        -\left(1+\frac{k_0}{2p}\right)^2\cA_L(K)\right]\sign(k_0)~~,
\eea
where the medium was taken to be parity invariant so that there is only 
one transverse polarization function.
There are contributions to $\Gamma_{\rm tot}$ both from creation and
annihilation of $\nu_R$ through several processes.  Depending on the
signs of $k_0$ and $p+k_0$ these processes can be divided into pair
creation/annihilation, Cherenkov radiation and scattering off charged
particles through virtual photon exchange \cite{Weldon83}.  
Calculating the imaginary part of the self-energy is thus a
convenient way of obtaining all processes that are allowed at the
one-loop level with a interacting photon correlation function.

As expected, this relaxation rate is identical to $\Gamma_{\rm
  gain}+\Gamma_{\rm loss}$ from \eq{gainloss} if we use a neutrino
distribution at zero chemical potential and if we express the field
correlators by virtue of \eq{BBEE2} in terms of $\cA_{T,L}$.

\Section{s:plasma}{Depolarization in a Relativistic Plasma}

In order to evaluate the depolarization rate explicitly for a specific
physical system we consider a relativistic QED plasma of the sort
encountered in the early universe where the chemical potentials of the
charged fermions are negligibly small. This system is characterized by
the temperature $T$ alone which is taken to be much larger than the
electron mass, but small enough that muons or pions are essentially
absent.

Even in such a relatively simple system, the lowest-order polarization
functions $\Pi_{T,L}$ depend on $K$ in very complicated ways. In order
to arrive at a first estimate we recall that Ref.~\cite{FY} implies
that the main contribution to the neutrino spin relaxation rate arises
from spin-flip scattering $\nu_L+e^\pm\to e^\pm+\nu_R$. The cross
section for this process has an infrared divergence which is
regularized by screening with a scale of order the photon plasma mass
$\cM=e T/3$.  
As a first estimate it is thus enough to obtain expressions for
$\Pi_{T,L}$ which are precise for $k_0,\,k\simleq eT$ for the leading
contribution.  In the high-temperature limit ($m_e\ll T$) they depend
only on the variable $k_0/k$. Explicitly one
finds~\cite{Klimov82}
\be{PiTL}
\ba{rcl}\displaystyle
   \Pi_T(K)&=&\displaystyle\frac{3\cM^2}{2}
        \left[\frac{k_0^2}{k^2}+(1-\frac{k_0^2}{k^2})
        \frac{k_0}{2k} \ln\left(\frac{k_0+k}{k_0-k}\right)\right]~~, \\[5mm]
   \Pi_L(K)&=&\displaystyle 3\cM^2
        (1-\frac{k_0^2}{k^2})
        \left[1-\frac{k_0}{2k}
        \ln\left(\frac{k_0+k}{k_0-k}\right)\right] ~~.
\ea
\ee
The real part of these functions is obtained by taking the modulus of
the argument of the logarithms. Below the light cone ($k_0^2<k^2$) 
the logarithms yield the imaginary parts
\be{PiTLim}
\ba{rcl}\displaystyle
   {\rm Im}\,\Pi_T(K)&=&\displaystyle-\frac{3\cM^2}{2}\pi
        (1-\frac{k_0^2}{k^2})\,\frac{k_0}{2k}\,\theta(k^2-k_0^2)~~, \\[5mm]
   {\rm Im}\,\Pi_L(K)&=&\displaystyle 3\cM^2\pi
        (1-\frac{k_0^2}{k^2})\,\frac{k_0}{2k}\,\theta(k^2-k_0^2)~~.
\ea
\ee
Above the light cone the only imaginary part of $\Pi_{T,L}$ comes 
from the $i\epsilon$ in the retarded prescription. Put another way,
above the light cone this approximation for $\Pi_{T,L}$ provides support
only on the discrete branches which correspond to transverse and
longitudinal plasmons. 

In the early universe, a typical thermal neutrino momentum is around
$3T$. Therefore, if indeed a typical momentum transfer is of order
$eT$ we are in a situation where the classical-trajectory approach
should be justified where it was assumed that $k_0,~k\ll p$ and
also $k_0\ll T$. Therefore, in \eq{GfIm} the infrared sensitive
Bose-Einstein term dominates which expands as 
$\sign(k_0)/(e^{\beta|k_0|}-1)=T/k_0$. Altogether we thus find for the
relaxation rate in the classical limit
\be{Gflargep} 
	\Gamma_{\rm tot}=
	\frac{\mu^2}{2\pi}\int_0^\infty dk\,k \int_{-k}^k
	dk_0\,\frac{T}{k_0}
	\frac{K^4}{k^2}\left[\cA_T(K)-\cA_L(K)\right]~~.
\ee 
If for the moment we ignore the resummation terms ${\rm Re}\Pi_{T,L}$
and ${\rm Im}\Pi_{T,L}$ in the denominator of \eq{ATL}, the classical
depolarization rate is found to be
\be{Gflargep2} 
	\Gamma_{\rm tot}=
	\alpha \mu^2 T^3\,\frac{2}{3}\int_0^\infty \frac{dk}{k}
\ee 
with $\alpha=e^2/4\pi\approx1/137$ the fine-structure constant.
Because this integral diverges 
it needs to be cut off by minimum and maximum
momentum transfers, 
leading to $\Gamma_{\rm tot}=\alpha\mu^2 T^3
(2/3) \ln(k_{\rm max}/k_{\rm min})$. Clearly, the infrared cutoff is
provided by the plasma mass $\cM=eT/3$ while $k_{\rm max}$ is given by
the neutrino momentum itself which is typically $3T$. With these
numbers we find $(2/3) \ln(k_{\rm max}/k_{\rm min})\approx 2$. Because
$\Gamma_{\rm tot}$ depends only logarithmically on the assumed
cutoffs we expect this simple estimate to be already rather precise.

We stress, however, that the initial assumption that only small
momentum transfers of order $eT$ were important was not justified. In
the classical approximation the momentum transfers are distributed as
$1/k$ so that the average is $\<k\>=(k_{\rm max}-k_{\rm min})/
\ln(k_{\rm max}/k_{\rm min})$. Therefore, with $k_{\rm max}\gg k_{\rm
min}$ the scale for a typical momentum transfer is set by $k_{\rm
max}$ and not by $k_{\rm min}$. Thus, even though the distribution of
momentum transfers peaks around $eT$, the average is still of order
$T$ and thus not small.

A numerically precise calculation of $\Gamma_{\rm tot}$ thus requires
the full quantum expressions. Moreover, while the approximate
expressions \eqs{PiTL}{PiTLim} for the polarization functions are
sufficient to implement the screening cutoff at small $k_0$ and $k$
and thus are sufficiently precise in the denominator of \eq{ATL}, they
are not guaranteed to be accurate enough in the numerator of
\eq{ATL}. While it is well known that the approximate expressions are
surprisingly accurate even for $|K|$ of order $T$, we still use the
full one-loop expressions which are given in
Ref.~\cite{EnqvistKS92}\footnote{In Eq.~(A.4) of
Ref.~\cite{EnqvistKS92} there is a factor $q^2/\omega^2$ missing in
the term below the light cone.}. The full one-loop expressions also
have the advantage of providing continuous support for $\cA_{T,L}$
above the light cone which corresponds to pair processes. Therefore,
we are able to compare directly the contribution to $\Gamma_{\rm tot}$
from pair processes (above the light cone) with that from spin-flip
transitions (below the light cone).

\begin{figure}
\unitlength=1mm
\begin{picture}(80,90)(0,0)
\includegraphics{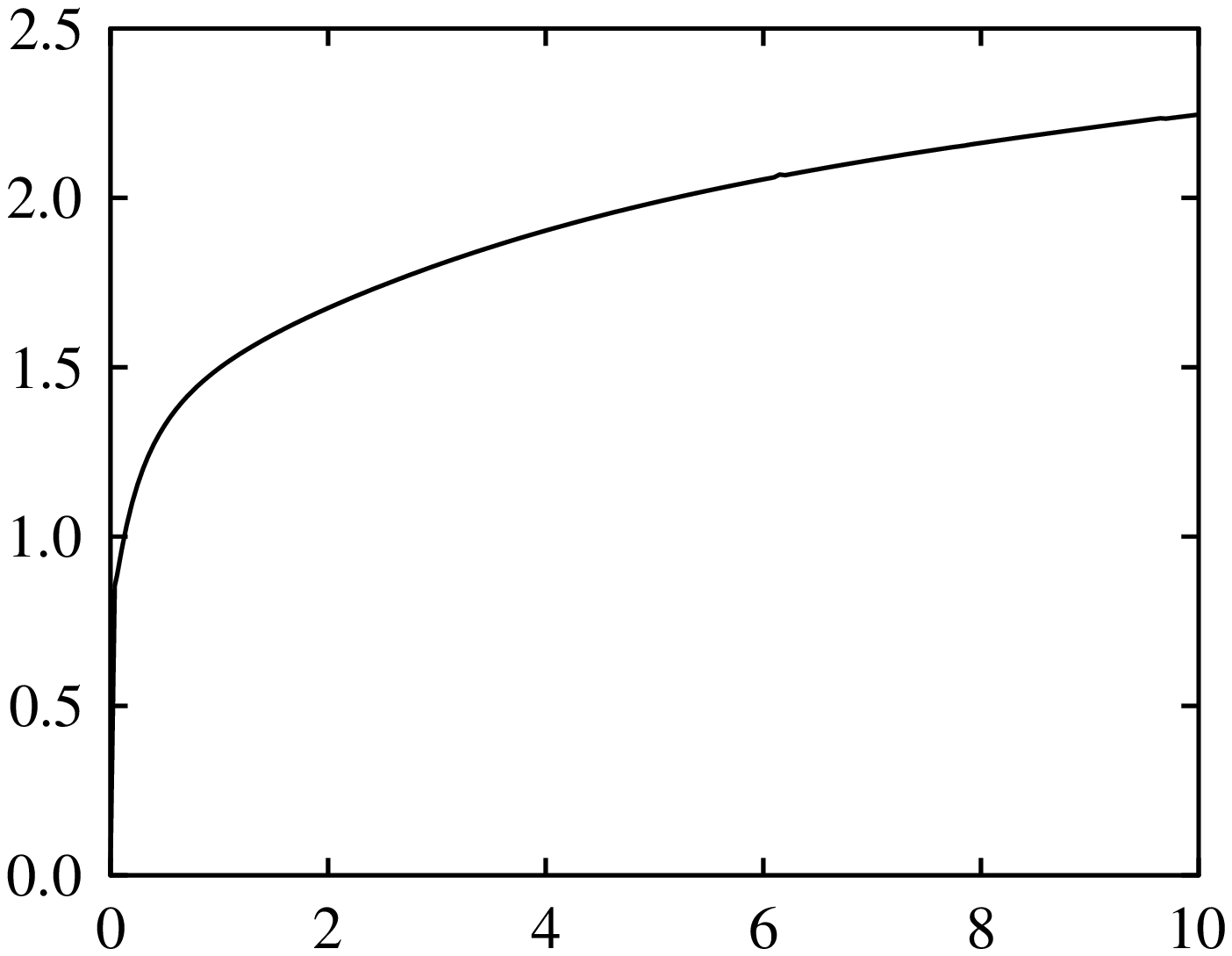}
   \put(0,0){}
   \put(20,45){\large$\frac{\Gamma_{\rm depol}}{\alpha\mu^2T^3}$}
   \put(85,0){\large$\frac{p}{T}$}
\end{picture}
   \caption{Depolarization rate as a function of neutrino momentum.
   \bort{Solid line: Spin-flip processes which are relevant for $K$ below
   the light cone. Dashed line: Pair processes which arise from $K$
   values above the light cone.}}
   \label{f:Gp}
\end{figure}

The numerical integration of the full quantum expression in \eq{GfIm}
yields the momentum dependent depolarization rate shown in \fig{f:Gp}\bort{,
where the solid curve is the contribution below the light cone, the
dashed curve from above}.  When we average these rates over a thermal
neutrino distribution we obtain
\be{Gfqnum}
        \<\Gamma_{\rm tot}\>=1.81 \alpha\mu^2 T^3~~,
\ee
where only $1.5 \%$ of the coefficient arise from pair processes. They
are indeed significantly subdominant as predicted in Ref.~\cite{FY}. 
Moreover, our simple estimate derived from the classical
approximation was surprisingly accurate. 

\Section{s:bigbang}{Early Universe}

\Subsection{ss:ssB}{Magnetic Moment Constraint in a Plasma}

The depolarization of the spin in the fluctuating electromagnetic
field of the early universe affects primordial nucleosynthesis.
Although there is a growing awareness that systematic errors are large
in the determination of the nucleosynthesis limit on the effective
number of neutrino species, it still seems reasonable to assume that
an extra neutrino degree of freedom is not allowed
\cite{nulimits}. Such an assumption then yields a cosmological limit
on the neutrino magnetic moment.

In order to avoid populating the right-handed component of our Dirac
neutrinos before BBN we need to require that $\Gamma_{\rm tot}$ is 
less than the Hubble rate at all times between the muon
annihilation and neutrino freeze-out epochs:
\be{GltH}
        \Gamma_{\rm tot}=0.0132\mu^2T^3<H=\frac{T^2}{m_{\rm Pl}}
        \left(\frac{4\pi^3g_*}{45}\right)^{1/2}~~.
\ee
Evidently, the most stringent bound comes from imposing this
constraint at as high a temperature as possible.
We take $g_*\simeq 10.75$ for the effective number of thermal degrees
of freedom which contribute to the energy density and thus
account for the electrons, 3 left-handed neutrino species, and the 
photons. With  $T=100$~MeV this leads to 
\be{mu}
        \mu< 6.2\times 10^{-11}\mu_B~~,
\ee
where $\mu_B=e/2m_e$ is the Bohr magneton.  This result puts
previous estimates \cite{FY}, where the infrared singularity in the
cross section of neutrino-electron scattering mediated by a
$t$-channel photon was estimated by a momentum cut-off at the Debye
mass, on a solid basis. 
Here we used the full resummed photon propagator to take into
account the screening effects correctly.

To obtain a more precise limit we should examine the relevant Boltzmann
equations. This is however not really warranted since the most
stringent limit on neutrino magnetic moments is still
obtained from energy loss considerations of helium burning globular
cluster stars. Plasmon decay would cool these stars too fast unless
$\mu \lsim 3\times 10^{-12}\mu_B$~\cite{georg,book}.  Of course, this
limit is valid only for neutrinos with mass less than a few keV.
There are additional bounds from SN~1987A \cite{Supernova,book} which
are valid up to the experimental $\nu_\tau$ mass limit of about
$24~\rm MeV$ and which are also more restrictive than the cosmological
limit, even though their exact numerical values differ significantly
between the works of different authors and are based on rather sparse
data. Still, unless these astrophysical limits are plagued with
implausibly large systematic errors we arrive at the conclusion that
neutrino dipole moments can affect nucleosynthesis only in connection
with a large-scale primordial magnetic field.

\Subsection{ss:lsB}{Large-scale Magnetic Field}

Another source for spin depolarization would be neutrino interactions
with a background magnetic field.  Let us assume for a moment that
such a hypothetical (constant) field exists and compare the neutrino
spin-flip rate due to a background field to spin flip rate induced by
the fluctuations of thermal photons.  This mechanism has been
considered previously by several authors
\cite{ElmforsGR96,EnqvistRS95,myyB} and it is of interest to compare
it with the depolarization in a stochastic electromagnetic field.  In
the classical picture the depolarization in a random field is a
consequence of the random walk that the polarization vector $\vP$
performs on the unit sphere, while in a constant background field the
polarization is attenuated by the helicity-measuring scattering that
projects the coherently rotating polarization vector onto the third
axis, corresponding to a damping of the off-diagonal elements in the density
matrix.

Though it has been derived several times before it 
easy enough to extract this damping directly from \eq{depeq}. We
can simply calculate the eigenvalues of $M$ and see that the damping of the
third component, in the limit of small $D$,  is
\be{lsBdamp}
        \Gamma_{b}=\frac{4\mu^2 B_\perp^2 D}
        {\omega_{\rm refr}^2+4\mu^2 B_\perp^2}~~.
\ee
At temperatures $1\MeV\le T\le 100 \MeV$ one finds
for effectively massless neutrinos \cite{NotzoldR88}
that $\omega_{\rm refr}=\xi\<p_0\>$ with
\be{xi}
   \xi=\frac{7\sqrt{2}\pi^2G_FT^4}{45m_Z^2}
        \left(1+C_\nu\frac{2m_Z^2}{m_W^2}\right)~~,
\ee
where $C_\nu=1$ for electron neutrinos so that
$\xi\simeq 1.1\EE{-20} ~(T/{\rm MeV})^4$, whereas
$C_\nu=0$ for
muon and tau neutrinos. The main contribution to $D$ comes
from neutrino elastic and inelastic scattering  with
leptons and equals half the total collision frequency
\cite{mckellar,EKT}.
Including scattering with electrons and neutrinos as well as annihilation
one finds at  temperature around $T=1\MeV$
\be{D}
    D=2.04\,G_{\rm F}^2 {T^5}~~,
\ee
for electron neutrinos.
Since from \eqs{xi}{D} we have that $\xi\<p_0\>/D\simeq 100$
the small $D$ approximation in \eq{lsBdamp} is good and would
remain so even if we include damping due to collisions with muons
in $D$.

Because the electrical conductivity of the universe is
large \cite{largecond},
a background magnetic field is imprinted on the
plasma and comoves with the expansion of the universe. Thus
flux conservation implies that the mean
rms field
scales with temperature like 
$\frac{3}{2}\<B^2_\perp\>^{1/2}=\<B^2\>^{1/2}=B_0(T/T_0)^2$.
We should however point out that
in the early universe a large conductivity implies a large
Reynolds number, and hence there is a possibility for turbulence which
can redistribute magnetic energy to various length scales. This was
verified in \cite{cascade}, where full magnetohydrodynamics
was simulated by using a simple MHD generalization of the cascade model
much used in studying hydrodynamical turbulence. In
the magnetic case it features
a transport of magnetic energy from small length scales to large
length scales. In a realistic situation the issue is then: has
the coherence length of the background field grown large enough
so it can be treated as a constant mean field on a given scale? Here we shall
just assume that such a mean field exists.
The limit from a large
scale background field concerns the combination
$\mu B_0$. Requiring that $\Gamma_b<H$ and using \eq{lsBdamp} we find that
\be{muB0}
        \mu B_0<\frac{T_0^2}{T^2}\left(\frac{3}{8}\right)^{1/2}
        \xi\<p_0\>\left(\frac{H}{D-H}\right)^{1/2}~~.
\ee
This constraint should be imposed at the temperature where the right hand
side is minimized (and $D>H$). With the damping rate in \eq{D} and $\<p_0\>=3T$
it happens at $T=1.5$ MeV which is above the kinetic freeze-out temperature
($2D\simgeq H$ gives $T\simgeq 1$ MeV), and we obtain
\be{muB0num}
 \left({\mu\over \mu_B}\right)\left({ B_0\over \MeV^2}\right)
<2.3\EE{-19}\, {T_0^2\over\MeV^2}~~.
\ee
Here we adopted $g_*=10.75$ in the Hubble rate.
We notice that this mechanism of depolarization in a large scale field is
most efficient at a rather low temperature, in contradistinction to the case
of a random field which gave a stricter constraint at high temperature.
The rate $\Gamma_b$
has been computed previously \cite{ElmforsGR96,EnqvistRS95} but was not
applied correctly at the neutrino freeze-out in \cite{EnqvistRS95} and
an approximate formula for $\Gamma_b$
was used in \cite{ElmforsGR96} leading to a somewhat too stringent bound.
Thus, the bound from a large scale field is better than the one from small
scale fluctuations if $B_0>3.7\EE{-9}$ (MeV)$^2$ 
at $T_0=1$ MeV, which is a
rather weak field compared to other typical scales.
The bound on a large scale magnetic
field from BBN is \cite{Blimits}
$B_0\lsim 4\times 10^{-5}\MeV^2$ at $T_0=0.01$ MeV.
(If the field is not homogeneous, then the limit is less stringent by
an order of magnitude.) There is thus room for
the large scale field to have been large enough to be the dominating
depolarization mechanism without contradicting BBN.
The existence or non-existence of such a primordial field should become
less speculative with the forthcoming experiments trying to measure
the strength of the intergalactic magnetic field \cite{faraday}.

\Section{s:summary}{Summary}

We have performed a detailed comparison of the classical and quantum
descriptions of neutrino depolarization by magnetic moment
interactions in the presence of a stochastic electromagnetic field.
Our main result is a general expression of the depolarization rate in
terms of the electric and magnetic field correlation functions. By
virtue of the fluctuation-dissipation theorem they are essentially
equivalent to the (imaginary parts) of the medium's dielectric
response functions.  Our analysis is exact up to second order in the
magnetic moment $\mu$ and to the order that the electromagnetic field
correlation functions are calculated.

We have evaluated the correlation functions and depolarization rate
explicitly for the case of a relativistic QED plasma. The neutrino
depolarization rate is dominated by spin-flip scattering on
relativistic electrons and positrons.  Even though the cross section
of this process peaks in the forward direction so that the
distribution of momentum transfers favors values of order $eT$, an
average momentum transfer is of order the neutrino
momentum. Therefore, the depolarization rate is not approximated well
by the classical description which is based on the assumption that a
typical momentum transfer is small relative to the neutrino
momentum. A numerically reliable result requires our full quantum
treatment.

Our precise calculation of the depolarization rate puts a previous
estimate \cite{FY} on a firm basis. Imposing the constraint that a right-handed
neutrino must not have been in equilibrium at nucleosynthesis we
derive the limit $\mu<6.2\EE{-11}\mu_B$ on the neutrino Dirac magnetic moment.
Other astrophysical limits are more restrictive by about an order of
magnitude, revealing that neutrino magnetic moments can affect
big-bang nucleosynthesis only in connection with large-scale
primordial magnetic fields which would provide for an additional
mechanism for left-right transitions.

\section*{Acknowledgments}

P.E.\ and K.E.\ thank the Aspen Center of Physics, where this work was
begun, for hospitality, a stimulating environment and financial
support.  We are also grateful to Avi Loeb and Leo Stodolsky for
discussions about the calculations in Ref.~\cite{LoebS89}, and to Kimmo
Kainulainen for useful remarks.  This work has been supported partly
by NorFA grant No. 96.15.053-O, EU contract CHRX CT930120, by the
Academy of Finland, and by the Deutsche Forschungsgemeinschaft under
grant SFB~375.


\end{document}